\documentclass[sigplan,nonacm]{acmart}
\makeatletter
\def\@ACM@checkaffil{
    \if@ACM@instpresent\else
    \ClassWarningNoLine{\@classname}{No institution present for an affiliation}%
    \fi
    \if@ACM@citypresent\else
    \ClassWarningNoLine{\@classname}{No city present for an affiliation}%
    \fi
    \if@ACM@countrypresent\else
        \ClassWarningNoLine{\@classname}{No country present for an affiliation}%
    \fi
}
\makeatother

\setcopyright{none}

\usepackage{centernot}
\usepackage{bussproofs}
\DeclareMathAlphabet{\mathcal}{OMS}{cmsy}{m}{n}
\SetMathAlphabet{\mathcal}{bold}{OMS}{cmsy}{b}{n}

\usepackage{pgfplots}
\pgfplotsset{compat=1.18}

\usepackage{xcolor}
\usepackage{soul}
\usepackage{mathtools}

\definecolor{Light}{gray}{.90}
\sethlcolor{Light}

\newcommand{\codett}[1]{\texttt{\hl{#1}}}
\DeclareMathOperator*{\largecomma}{\mathop{\vcenter{\hbox{\Huge\texttt{,}}}}}

\usepackage{wrapfig}

\begin{document}

  \title{A Word Sampler for Well-Typed Functions}
  \begin{abstract}
    We describe an exact sampler for a simply-typed, first-order functional programming language. Given an acyclic finite automaton, $\alpha_{\varnothing}$, it samples a random function uniformly without replacement from well-typed functions in $\mathcal{L}(\alpha_{\varnothing})$. This is achieved via a fixed-parameter tractable reduction from a syntax-directed type system to a context-free grammar, preserving type soundness and completeness w.r.t. $\mathcal{L}(\alpha_{\varnothing})$, while retaining the robust metatheory of formal languages.
  \end{abstract}

  \author{Breandan Considine}
  \email{bre@ndan.co}

  \maketitle

  \section{Introduction}

Consider a simply-typed language with the following terms:
  \newcommand{\nt}[1]{\texttt{#1}}      
  \newcommand{\tok}[1]{\codett{#1}}     
  \newcommand{\gap}{\;\;}                 
  \[
    \hspace{-0.1cm}\begin{array}{ccl}
      \nt{FUN} & ::= & \tok{fun}\gap \tok{f0}\gap \tok{(}\gap \nt{PRM}\gap \tok{)}\gap \tok{:}\gap \mathbb{T} \gap \tok{=}\gap \nt{EXP} \\
      \nt{PRM} & ::= & \nt{PID}\gap \tok{:}\gap \mathbb{T} \;\mid\; \nt{PRM}\gap \tok{,}\gap \nt{PID}\gap \tok{:}\gap \mathbb{T} \\
      \nt{EXP} & ::= & \ulcorner\mathbb{N}\lrcorner \;\mid\; \ulcorner\mathbb{B}\lrcorner \;\mid\; \nt{PID} \;\mid\; \nt{INV} \;\mid\; \nt{IFE} \;\mid\; \nt{OPX}\\
      \nt{OPX} & ::= & \codett{(}\gap\nt{EXP}\gap \nt{OPR}\gap \nt{EXP}\gap\codett{)} \\
      \nt{IFE} & ::= & \tok{if}\gap \nt{EXP}\gap \tok{\{}\gap \nt{EXP}\gap \tok{\}}\gap
      \tok{else}\gap \tok{\{}\gap \nt{EXP}\gap \tok{\}} \\
      \nt{INV} & ::= & \nt{FID}\gap \tok{(}\gap \nt{ARG}\gap \tok{)} \\
      \nt{ARG} & ::= & \nt{EXP} \;\mid\; \nt{ARG}\gap \tok{,}\gap \nt{EXP} \\
      \nt{OPR} & ::= & \tok{+} \;\mid\; \tok{*} \;\mid\; \tok{<} \;\mid\; \tok{==} \\
      \nt{PID} & ::= & \tok{p1} \;\mid\; \ldots \;\mid\; \tok{pk} \\
      \nt{FID} & ::= & \tok{f0} \;\mid\; \tok{f1} \;\mid\; \ldots \;\mid\; \tok{fn}\\
      \ulcorner\mathbb{B}\lrcorner & ::= & \tok{true} \;\mid\; \tok{false}\\
      \ulcorner\mathbb{N}\lrcorner & ::= & \tok{1} \;\mid\; \tok{2} \;\mid\; \tok{3} \;\mid\; \ldots
    \end{array}
  \]

  \noindent At the type level, we will assume an ambient global context, $\Gamma$, consisting of invokable named functions, and a finite type universe with two primitive types, $\mathbb{B}$ and $\mathbb{N}$.
  \[
    \begin{array}{rcl}
      \Gamma & ::= & \varnothing \texttt{ }\mid \texttt{ }\Gamma, \codett{f\_} : (\tau_{1}, \ldots, \tau_{k}) \rightarrow \tau\\
      \mathbb{T} & ::= & \hspace{0.039cm}\mathbb{B}\texttt{ } \mid \texttt{ }\mathbb{N}\texttt{ } \mid \texttt{ }\tau^{(3)}\texttt{ } \mid \texttt{ }\ldots\texttt{ }\mid \texttt{ }\tau^{(d)}
    \end{array}
  \]

  Let us define a fragment of the typing judgements for \texttt{IFE}, \texttt{INV}, and \texttt{OPX}, which are mostly conventional.

  \begin{prooftree}
    \AxiomC{$\Gamma \vdash e_c: \mathbb{B}$}
    \AxiomC{$\Gamma \vdash e_\top: \tau$}
    \AxiomC{$\Gamma \vdash e_\bot: \tau$}
    \RightLabel{$\texttt{IFE}$}
    \TrinaryInfC{$\Gamma \vdash \tok{if}\gap e_c \gap \tok{\{}\gap e_\top \gap \tok{\}}\gap\tok{else}\gap\tok{\{}\gap e_\bot \gap\tok{\}}: \tau$}
    \DisplayProof
    \vskip 0.3cm
    \AxiomC{$\Gamma \vdash \tok{f\_}: (\tau_1, \ldots, \tau_m)\rightarrow\tau$}
    \AxiomC{$\Gamma \vdash e_i:\tau_i \:\:\forall i \in [1, m]$}
    \RightLabel{$\texttt{INV}$}
    \BinaryInfC{$\Gamma \vdash \tok{f\_}\gap\tok{(}\gap e_1 \gap\tok{,}\gap \ldots \gap\tok{,}\gap e_m\gap\tok{)}: \tau$}
    \DisplayProof
    \vskip 0.3cm
    \AxiomC{$\delta_{\texttt{OPR}}(\odot, \tau, \tau') = \hat\tau$}
    \AxiomC{$\Gamma \vdash e_1 :\tau$}
    \AxiomC{$\Gamma \vdash e_2 :\tau'$}
    \RightLabel{$\texttt{OPX}$}
    \TrinaryInfC{$\Gamma \vdash \tok{(}\gap e_1 \odot e_2 \gap\tok{)}: \hat\tau$}
\end{prooftree}\vspace{0.2cm}

  \noindent where $\delta_{\texttt{OPR}}: \Sigma_{\texttt{OPR}} \times \mathbb{T} \times \mathbb{T} \rightharpoonup \mathbb{T}$ is defined as follows:
 \begin{equation*}
   \delta_{\texttt{OPR}}(\odot, \tau, \tau') = \begin{cases}
     \mathbb{B} & \text{ if } \odot \in \{\codett{<}\}, \tau, \tau' :\mathbb{N}\\
     \mathbb{N} & \text{ if } \odot \in \{\codett{+}, \codett{*}\}, \tau, \tau' :\mathbb{N}\\
     \mathbb{B} & \text{ if } \odot \in \{\codett{==}\}, \tau = \tau'\:\: \forall \tau, \tau': \mathbb{T}
   \end{cases}
 \end{equation*}

  \noindent We will encode the type checker as a context-free grammar.

  \section{Notation}

  Recall that a context-free grammar (CFG) is a quadruple, $\langle \Sigma, V, P, S\rangle$, consisting of terminals $(\Sigma)$, nonterminals $(V)$, productions $\big(P \subset V \times (V\cup \Sigma)^*\big),$ and a start symbol $(S)$. Also, a finite automaton (FA) is a quintuple $\langle Q, \Sigma, \delta, q_\alpha, F\rangle$, with states $(Q)$, an alphabet $(\Sigma)$, transitions $(\delta \subseteq Q \times \Sigma \times Q)$, an initial state $(q_\alpha)$, and accepting states $(F \subseteq Q)$. These devices generate words in languages, denoted $\mathcal{L}(\cdot) \subseteq \Sigma^*$, that are context-free and regular, respectively.

  A few notational rules for CFG compilation will be helpful:

  \begin{small}
  \begin{prooftree}
    \AxiomC{$\codett{.}$}
    \RightLabel{$\Sigma$}
    \UnaryInfC{$\codett{.} \in \Sigma$}
    \DisplayProof
    \hskip 1em
    \AxiomC{$(\sigma_0 \rightarrow \sigma_{1..n}) \in P$}
    \RightLabel{$P_V$}
    \UnaryInfC{$\bigcup_{i=0}^n\{\sigma_i\} \setminus \Sigma \in V$}
    \DisplayProof
    \hskip 1em
    \AxiomC{$(\sigma_0 \rightarrow \sigma_{1..n}) \in P$}
    \RightLabel{$P_\Sigma$}
    \UnaryInfC{$\bigcup_{i=1}^n \{\sigma_i\} \setminus V \in \Sigma$}
  \end{prooftree}
  \end{small}

  \newcommand{\cjoin}[2]{\mathop{\largecomma}\limits_{#1}^{#2}}

  \noindent The notation $\cjoin{}{} (\cdot)$ is a macro for a comma-separated list.\footnote{i.e., $\cjoin{i=1}{m}(x_i)\;:=\; x_1 \gap \codett{,} \gap \ldots \gap x_m \text{ if } m > 1 \text{ else } x_1 \text{ if } m = 1 \text{ else } \varepsilon$.}


  \section{Method}

  We want to permit functions of up to arity-$k$, so the start symbol, $S_\Gamma$, will need to express each of these possibilities:

  \begin{tiny}
  \begin{prooftree}
    \AxiomC{$\langle\vec\tau,\dot\tau\rangle \in \mathbb{T}^{0..k}\times\mathbb{T}$}
    \AxiomC{$\vec\tau_{0..|\vec\tau|} \in \vec\tau$}
    \RightLabel{$\texttt{FUN}_{\varphi}$}
    \BinaryInfC{$\Big(S_\Gamma \rightarrow
    \tok{fun}\gap \tok{f0}\gap
    \tok{(}\cjoin{i=1}{|\vec\tau|}\big(p_i \gap\tok{:}\gap \vec\tau_{i}\big)\:\tok{)}\gap
    \tok{:}\gap \dot\tau \gap \tok{=}\gap \texttt{EXP}[\dot\tau, \vec\tau \rightarrow \dot\tau]
    \Big)\in P_\Gamma$}
  \end{prooftree}
  \end{tiny}
  We will decorate $\texttt{EXP}$ nonterminals with a pair, \texttt{EXP}$[\cdot, \cdot]$, of (1) the expression's local return type $(\tau)$, and (2) available parameters $(\vec\tau)$ and expected return type $(\dot\tau)$ for $\codett{f0}:\vec\tau\rightarrow \dot\tau$:
  \begin{tiny}
  \begin{prooftree}
    \AxiomC{$\texttt{EXP}[\tau, \vec\tau \rightarrow \dot\tau] \in V_\Gamma$}
    \AxiomC{$\Gamma \vdash \codett{f\_} : (\tau_{1}, \ldots, \tau_{m}) \rightarrow \tau$}
    \RightLabel{$\texttt{INV}_{\varphi}$}
    \BinaryInfC{$\big(\nt{EXP}[\tau,\vec\tau \rightarrow \dot\tau] \;\rightarrow\; \tok{f\_}\gap \tok{(} \cjoin{i=1}{m}\nt{EXP}[\tau_{i},\vec\tau \rightarrow \dot\tau] \:\tok{)}\big)\in P_\Gamma$}
    \DisplayProof
    \vskip 1.5em
    \AxiomC{$\texttt{EXP}[\tau, \vec\tau \rightarrow \dot\tau] \in V_\Gamma$}
    \AxiomC{$\tau = \dot\tau$}
    \AxiomC{$\vec\tau_{0..|\vec\tau|} \in \vec\tau$}
    \RightLabel{$\texttt{REC}_{\varphi}$}
    \TrinaryInfC{$\big(\nt{EXP}[\tau,\vec\tau \rightarrow \dot\tau] \;\rightarrow\; \tok{f0}\gap \tok{(} \cjoin{i=1}{|\vec\tau|}\nt{EXP}[\vec\tau_{i},\vec\tau \rightarrow \dot\tau] \:\tok{)}\big)\in P_\Gamma$}
    \DisplayProof
    \vskip 1.5em
    \AxiomC{$\texttt{EXP}[\tau, \vec\tau \rightarrow \dot\tau] \in V_\Gamma$}
    \AxiomC{$\tau = \tau'$}
    \AxiomC{$\tau, \tau' \in \mathbb{T}$}
    \RightLabel{$\texttt{IFE}_{\varphi}$}
    \TrinaryInfC{$
    \left(
      \begin{array}{l}
        \nt{EXP}[\tau,\vec\tau \rightarrow \dot\tau] \;\rightarrow\;
        \tok{if}\gap \nt{EXP}[\mathbb{B},\vec\tau \rightarrow \dot\tau]\gap \tok{\{}\gap \nt{EXP}[\tau,\vec\tau \rightarrow \dot\tau]\gap \tok{\}}\\[0.15cm]
        \phantom{\nt{EXP}[\tau,\vec\tau \rightarrow \dot\tau] \;\rightarrow\;}\;\tok{else}\gap \tok{\{}\gap \nt{EXP}[\tau',\vec\tau \rightarrow \dot\tau]\gap \tok{\}}
      \end{array}
    \right)\in P_\Gamma
    $}
    \DisplayProof
    \vskip 1.5em
    \AxiomC{$\texttt{EXP}[\hat\tau, \vec\tau \rightarrow \dot\tau] \in V_\Gamma$}
    \AxiomC{$\delta_{\texttt{OPR}}(\odot,\tau,\tau')=\hat\tau$}
    \AxiomC{$\odot\in \{\codett{==}, \codett{<}, \codett{+}, \codett{*}\}$}
    \RightLabel{$\texttt{OPX}_{\varphi}$}
    \TrinaryInfC{$\big(\nt{EXP}[\hat\tau,\vec\tau \rightarrow \dot\tau] \;\rightarrow\; \codett{(}\gap\nt{EXP}[\tau,\vec\tau \rightarrow \dot\tau] \;\odot\; \nt{EXP}[\tau',\vec\tau \rightarrow \dot\tau]\gap\codett{)}\big)\in P_\Gamma$}
    \DisplayProof
    \vskip 1.5em
    \AxiomC{$\texttt{EXP}[\tau, \vec\tau \rightarrow \dot\tau] \in V_\Gamma \gap\gap\exists\vec\tau_{i}=\tau$}
    \RightLabel{$\texttt{PID}_{\varphi}$}
    \UnaryInfC{$\big(\nt{EXP}[\tau,\vec\tau \rightarrow \dot\tau] \rightarrow \tok{pi}\big)\in P_\Gamma$}
    \DisplayProof
    \hskip 0.5em
    \AxiomC{$\texttt{EXP}[\tau, \vec\tau \rightarrow \dot\tau] \in V_\Gamma\:\:\:\tau \in \{\mathbb{B}, \mathbb{N}\}\:\:\:\codett{\_}:\tau$}
    \RightLabel{$\texttt{LIT}_{\varphi}$}
    \UnaryInfC{$(\texttt{EXP}[\tau, \vec\tau \rightarrow \dot\tau] \rightarrow \tok{\_})\in P_\Gamma$}
  \end{prooftree}
  \end{tiny}

  \vspace{0.2cm}

  The resulting grammar, $G_\Gamma:=\langle \Sigma, V_\Gamma, P_\Gamma, S_\Gamma\rangle$, will be put into Chomsky Normal Form (CNF), $G_\Gamma'$, pruning productions containing unreachable or unproductive nonterminals and refactoring each production to either $(w \rightarrow x z): V\times V^2$ or $(w \rightarrow t): V\times\Sigma$. During normalization, arbitrary CFGs may undergo a quadratic blowup in space~\cite{lange2009cnf}, however, as this grammar does not use $\varepsilon$ or contain unary production chains, we can approximate the enlargement as being linear in $|G_\Gamma|$.

  \subsection{Space complexity}

  Let us attempt to estimate $|G_\Gamma'|$ in terms of the contribution from each constructor. Following the convention of Lange and Lei{\ss}~\cite{lange2009cnf}, we define $|G| = \sum_{w\in V}\sum_{w\rightarrow \sigma} |w\sigma|$. We will also use $|\cdot|'$ to denote binarized production size, which depends on the specific binarization technique, but is bounded by:
    \begin{align*}
      |w\rightarrow \sigma|': P \rightarrow \mathbb{N}
      \begin{cases}
         = |w\sigma| &\text{ if } |w\sigma| \leq 3\\
         \leq 3|w \sigma| - 1 &\text{ otherwise.}
      \end{cases}
    \end{align*}

  The leading term clearly depends on $\texttt{FUN}_\varphi$, which generates function signatures up to arity-$k$, with $d=|\mathbb{T}|$ types. If one considers permuted orderings of the input type signature, $\vec\tau$, as identical, its cost improves to $d\sum_{i = 0}^{k}{d+i - 1 \choose i-1}$, however we will adhere to the na\"ive interpretation, which takes an arithmetico-geometric form, $\sum_{p=1}^{k}pd^{p+1}$, whose Lange-Lei{\ss} size is primarily determined by three factors:\vspace{-0.1em}
  \begin{footnotesize}
  \begin{align*}
    |\texttt{FUN}_\varphi|' &= \big|S_\Gamma \rightarrow \tok{fun}\ \tok{f0}\ \tok{(}\cjoin{i=1}{|\vec\tau|}p_i\ \tok{:}\ \vec\tau_{i}\:\tok{)}\ \tok{:}\ \tau\ \tok{=}\ \texttt{EXP}[\tau, \vec\tau\rightarrow\tau]\big|'\leq\boxed{12|\vec\tau|+23}\\
    |\texttt{REC}_\varphi|' &= \big|\texttt{EXP}[\tau,\vec\tau \rightarrow \tau] \rightarrow \tok{f0}\ \tok{(}\cjoin{i=1}{|\vec\tau|}\texttt{EXP}[\vec\tau_{i}, \vec\tau\rightarrow\tau]\:\tok{)}\big|'\leq\boxed{6|\vec\tau| + 8}\\
    |\texttt{PID}_\varphi|' &= |\vec\tau|\cdot\big|\texttt{EXP}[\vec\tau_{i},\vec\tau \rightarrow \tau] \rightarrow \tok{pi}\big|' = \boxed{2|\vec\tau|}
  \end{align*}
  \end{footnotesize}
  Letting $p=|\vec\tau|$ and assembling these factors, we have,
  \begin{align*}
    |G_\Gamma'| \simeq &\sum_{p=0}^k d^{p+1}\big(\underbrace{(12p+23)}_{|\nt{FUN}_\varphi|'} + \underbrace{(6p+8)}_{|\nt{REC}_\varphi|'} \big) + \sum_{p=1}^k d^{p+1}\underbrace{(2p)}_{\mathclap{|\nt{PID}_\varphi|'}}\\
    \simeq &\sum_{p=1}^{k}(20p + 31)d^{p+1} \simeq  \frac{20kd^{k+2}}{(d-1)^2}+\mathcal{O}\left(\frac{d^{k+2}}{d-1}\right)+\ldots
  \end{align*}
  and lower-degree terms. While our analysis omits $\texttt{INV}_\varphi$, et al., their contributions are less sensitive to the parameter $k$.

  \subsection{Sampling}

  The context-free languages have the pleasant property of being closed under intersection with regular languages~\cite{bar1961formal}, with an explicit construction given by Salomaa~\cite{salomaa1973formal}. In brief, for every production $\text{W} \rightarrow \text{X Z}$ in the CNF grammar and state triple $p, q, r: Q$ in the automaton one creates synthetic (1) binary productions $p\text{W}r \rightarrow p\text{X}q\:\:\:q\text{Z}r$, (2) unit productions $p\text{W}q \rightarrow a$ for every $\text{W} \rightarrow a$ and $p, q$ such that $\delta(p, a) = q$, and (3) start productions $\text{S} \rightarrow q_\alpha\text{S}q_\omega$ for every final state $q_\omega$.

  Once constructed, a variety of methods for enumerating and sampling CFGs can be applied (e.g.,~\cite{considine2024tree, chiang2007hierarchical, piantadosi2023enumerate}). We use Considine's~\cite{considine2024tree}, which supports sampling words in language intersections parameterized by an acyclic FA, in which case the intersection will be representable as an acyclic FA, $\alpha_{\cap}$.

  This representation can be derminimized to produce an acyclic deterministic FA and then decoded left-to-right using an autoregressive model, or, if uniformity over $\ell_\cap$ is desired, by constructing a bijection, $b: \mathbb{Z}_{|\ell_\cap|}\leftrightarrow \ell_\cap$, and then drawing samples from a pseudorandom source (e.g., a linear feedback shift register). The latter method has the virtue of perfect parallelizability, although we evaluate this method serially.

  \section{Evaluation}

  We evaluate the sampler for small arity, $k\in [1,3]$, with fixed $|\Gamma|=18, |\mathbb{T}|=7$ (see Appendix~\ref{sec:ac}). This generates tractable CNF grammars, $|G_\Gamma'| \in [1.9\times 10^4, 9.9\times 10^5]$, from which we then sample words on an Apple M4 with 16 GB of memory.

  First, we sample words from a slice, $\sigma \leftsquigarrow \mathcal{L}(G_\Gamma') \cap \Sigma^n$, and measure total time to first sample ($\mu$ TTFS). Next, using our dataset of random functions obtained during slice sampling, we will replace $(\tok{:}\:\tau\:\tok{=})$ with a hole $(\tok{:}\: \Sigma\:\tok{=})$ and resample $\sigma' \leftsquigarrow \mathcal{L}(G_\Gamma') \cap (\ldots\:\tok{:}\:\Sigma\:\tok{=}\:\ldots)$, which we call type inference.\vspace{-0.2cm}

  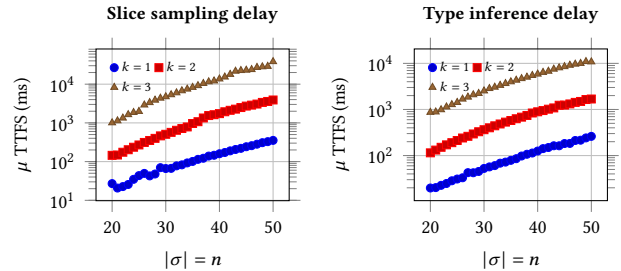
\begin{figure}[H]
    \centering
    \begin{minipage}{0.49\columnwidth}
      \centering
      \begin{tikzpicture}
        \begin{axis}[
        width=\linewidth, height=3.6cm,
        xlabel={$|\sigma|=n$}, ylabel={$\mu$ TTFS (ms)},
        title={\scriptsize{\textbf{Slice sampling delay}}},
        ymode=log,
        xmajorgrids, ymajorgrids,
        tick align=outside,
        tick label style={font=\scriptsize},
        label style={font=\scriptsize},
        legend columns=2,
        legend style={draw=none, fill=none, font=\scriptsize, at={(0.03,0.98)}, anchor=north west, nodes={scale=0.7, transform shape}},
        scaled y ticks=false,
        ]
        \addplot+[only marks, mark=*, mark size=1.5pt]
        table[col sep=comma, x=n, y=mean_ttfs_ms]{
          n,mean_ttfs_ms
          20,27.05
          21,20.75
          22,22.65
          23,25.65
          24,34.85
          25,43.5
          26,49.6
          27,42.75
          28,47.85
          29,69.75
          30,66.25
          31,66.9
          32,77.65
          33,82.4
          34,92.5
          35,101.0
          36,114.3
          37,122.15
          38,138.9
          39,145.8
          40,159.55
          41,171.85
          42,189.55
          43,205.7
          44,220.0
          45,242.15
          46,257.8
          47,281.75
          48,308.35
          49,327.75
          50,353.4
        };
        \addplot+[only marks, mark=square*, mark size=1.5pt]
        table[col sep=comma, x=n, y=mean_ttfs_ms]{
          n,mean_ttfs_ms
          20,145.25
          21,147.9
          22,174.6
          23,202.6
          24,235.15
          25,269.35
          26,312.8
          27,351.45
          28,398.8
          29,461.9
          30,507.95
          31,569.5
          32,644.9
          33,715.6
          34,786.05
          35,963.7
          36,1073.4
          37,1339.9
          38,1546.9
          39,1627.15
          40,1734.3
          41,1938.4
          42,2115.55
          43,2267.0
          44,2490.35
          45,2717.05
          46,2878.45
          47,3065.9
          48,3383.15
          49,3580.85
          50,3907.65
        };
        \addplot+[only marks, mark=triangle*, mark size=1.5pt]
        table[col sep=comma, x=n, y=mean_ttfs_ms]{
          n,mean_ttfs_ms
          20,1005.5
          21,1132.15
          22,1333.05
          23,1633.95
          24,1856.05
          25,1978.3
          26,2900.25
          27,3344.15
          28,3864.75
          29,4355.9
          30,4744.05
          31,5341.85
          32,5986.6
          33,6762.95
          34,7468.8
          35,8270.4
          36,9174.45
          37,10416.85
          38,11540.75
          39,12119.55
          40,13643.15
          41,15313.7
          42,17913.25
          43,21658.65
          44,22495.4
          45,22996.15
          46,23839.55
          47,27062.25
          48,28112.65
          49,28886.45
          50,38244.9
        };
        \legend{$k=1$,$k=2$,$k=3$}
        \end{axis}
      \end{tikzpicture}
    \end{minipage}\hfill
    \begin{minipage}{0.49\columnwidth}
      \centering
      \begin{tikzpicture}
        \begin{axis}[
        title={\scriptsize{\textbf{Type inference delay}}},
        width=\linewidth, height=3.6cm,
        xlabel={$|\sigma|=n$}, ylabel={$\mu$ TTFS (ms)},
        ymode=log,
        xmajorgrids, ymajorgrids,
        tick align=outside,
        tick label style={font=\scriptsize},
        label style={font=\scriptsize},
        legend columns=2,
        legend style={draw=none, fill=none, font=\scriptsize, at={(0.03,0.98)}, anchor=north west, nodes={scale=0.7, transform shape}},
        scaled y ticks=false,
        ]
        \addplot+[only marks, mark=*, mark size=1.5pt]
        table[col sep=comma, x=n, y=mean_ttfs_ms]{
          n,mean_ttfs_ms
          20,19.8
          21,20.2
          22,22.6
          23,24.8
          24,28.4
          25,31.0
          26,33.0
          27,42.6
          28,42.4
          29,45.2
          30,52.6
          31,56.8
          32,60.6
          33,68.4
          34,72.2
          35,78.8
          36,85.6
          37,97.8
          38,106.8
          39,113.6
          40,125.4
          41,141.6
          42,149.6
          43,163.2
          44,163.0
          45,182.4
          46,183.4
          47,212.4
          48,215.6
          49,241.6
          50,261.0
        };
        \addplot+[only marks, mark=square*, mark size=1.5pt]
        table[col sep=comma, x=n, y=mean_ttfs_ms]{
          n,mean_ttfs_ms
          20,114.6
          21,131.6
          22,151.2
          23,170.8
          24,190.4
          25,212.4
          26,240.6
          27,261.8
          28,294.4
          29,328.8
          30,363.8
          31,400.4
          32,442.2
          33,484.6
          34,532.0
          35,576.6
          36,629.0
          37,681.6
          38,741.4
          39,838.4
          40,890.2
          41,946.4
          42,1014.4
          43,1080.6
          44,1235.0
          45,1245.4
          46,1318.8
          47,1424.8
          48,1493.6
          49,1642.8
          50,1686.4
        };
        \addplot+[only marks, mark=triangle*, mark size=1.5pt]
        table[col sep=comma, x=n, y=mean_ttfs_ms]{
          n,mean_ttfs_ms
          20,867.6
          21,883.2
          22,987.2
          23,1135.2
          24,1274.2
          25,1419.0
          26,1680.2
          27,1854.4
          28,2084.0
          29,2224.2
          30,2526.4
          31,2844.2
          32,3044.8
          33,3335.8
          34,3650.4
          35,3984.8
          36,4317.0
          37,4671.4
          38,5019.6
          39,5462.2
          40,5855.2
          41,6366.6
          42,6754.2
          43,7491.6
          44,8011.2
          45,8494.2
          46,9104.2
          47,9618.2
          48,9974.4
          49,10931.6
          50,10779.6
        };
        \legend{$k=1$,$k=2$,$k=3$}
        \end{axis}
      \end{tikzpicture}
    \end{minipage}
    \vspace{-0.9em}
    \caption{Slice sampling and type inference delay (log-scaled) vs. sequence length $|\sigma|=n$ for arities $k\in [1,3]$.}
    \vspace{-0.6em}
  \end{figure}

  Once the first sample is obtained, we observe bounded delay of $1786 \pm 817 $ ns $(\mu \pm \sigma)$, or an average throughput of $\sim 5.6\times 10^5$ samples per second. Enumeration delay does not appear strongly correlated with either arity or word length.

  \section{Related work}

  Prior work demonstrates how to embed a deterministic CFL into a type-system~\cite{roth2021study}, but the reverse direction remains largely unexplored. Existing work on constrained decoding (e.g., Willard et al.~\cite{willard2023efficient}) shows that syntactic soundness is feasible to guarantee, but the sample space is often ill-defined or an overapproximation to the space of semantically valid candidates. Frank et al.~\cite{frank2024generating} introduce a type-theoretic method for sampling well-typed terms, but their method does not guarantee statistical uniformity or syntactic completeness. Finally, Bendkowski~\cite{bendkowski2016boltzmann} uses techniques from enumerative combinatorics to sample closed $\lambda$-terms of the simply-typed variety, which is most closely related to this line of work.

  \section{Conclusion}

  We have presented a CFG embedding and exact sampler for well-typed functions, tractable for small-$k$. Extensions to straight-line programs, higher-order functions, and richer typing formalisms such as subtyping, parametric polymorphism, and substructural constraints are conceivable. Due to the cost of materializing $G_\Gamma'$, it would be advantageous to construct the constituent productions lazily, as only a small fraction may participate in a given language intersection. Another direction would be to collapse syntactic symmetries by quotienting productions, e.g., by semantic invariants or $\alpha$-equivalence. We leave these possibilities for future work.

  \clearpage
  \section{Acknowledgements}

  This paper is dedicated to Ori Roth, whose work on typelevel parsing inspired the author to pursue this direction.

  \bibliography{acmart}

@article{bar1961formal,
    title        = {On formal properties of simple phrase structure grammars},
    author       = {Bar-Hillel, Yehoshua and Perles, Micha and Shamir, Eli},
    year         = 1961,
    journal      = {Sprachtypologie und Universalienforschung},
    publisher    = {Akademie-Verlag},
    volume       = 14,
    pages        = {143--172}
}

@article{roth2021study,
    title        = {Study of the Subtyping Machine of Nominal Subtyping with Variance},
    author       = {Roth, Ori},
    year         = 2021,
    journal      = {arXiv preprint arXiv:2109.03950},
    url          = {https://arxiv.org/pdf/2109.03950.pdf}
}

@misc{piantadosi2023enumerate,
    title        = {How to enumerate trees from a context-free grammar},
    author       = {Steven T. Piantadosi},
    year         = 2023,
    eprint       = {2305.00522},
    archiveprefix = {arXiv},
    primaryclass = {cs.CL}
}

@article{chiang2007hierarchical,
    title={Hierarchical phrase-based translation},
    author={Chiang, David},
    journal={Computational Linguistics},
    volume={33},
    number={2},
    pages={201--228},
    year={2007},
    publisher={MIT Press One Rogers Street, Cambridge, MA 02142-1209, USA journals-info~…}
}

@article{willard2023efficient,
    title={Efficient guided generation for {LLMs}},
    author={Willard, Brandon T and Louf, R{\'e}mi},
    journal={arXiv preprint arXiv:2307.09702},
    year={2023}
}

@article{considine2024tree,
    title={A Tree Sampler for Bounded Context-Free Languages},
    author={Considine, Breandan},
    journal={arXiv preprint arXiv:2408.01849},
    year={2024}
}

@inproceedings{bendkowski2016boltzmann,
    title={Boltzmann samplers for closed simply-typed lambda terms},
    author={Bendkowski, Maciej and Grygiel, Katarzyna and Tarau, Paul},
    booktitle={International Symposium on Practical Aspects of Declarative Languages},
    pages={120--135},
    year={2016},
    organization={Springer}
}

@book{salomaa1973formal,
    author = {Salomaa, Arto},
    title = {Formal languages},
    publisher = {Academic Press},
    year = {1973},
    pages = {59--61},
    address = {New York}
}

@article{frank2024generating,
    title={Generating Well-Typed Terms That Are Not “Useless”},
    author={Frank, Justin and Quiring, Benjamin and Lampropoulos, Leonidas},
    journal={Proceedings of the ACM on Programming Languages},
    volume={8},
    number={POPL},
    pages={2318--2339},
    year={2024},
    publisher={ACM New York, NY, USA}
}

@article{lange2009cnf,
    title={To CNF or not to CNF? An efficient yet presentable version of the CYK algorithm},
    author={Lange, Martin and Lei{\ss}, Hans},
    journal={Informatica Didactica},
    volume={8},
    number={2009},
    pages={1--21},
    year={2009}
}
  \appendix
  \section{Ambient context ($\Gamma$)}\label{sec:ac}
  \begin{align*}
  \phantom{....}\texttt{i2s} & :       \texttt{Int} \rightarrow \texttt{Str},\\
  \phantom{....}\texttt{s2i} & :       \texttt{Str} \rightarrow \texttt{Int},\\
  \phantom{....}\texttt{i2f} & :       \texttt{Int} \rightarrow \texttt{Float},\\
  \phantom{....}\texttt{len} & :       \texttt{Str} \rightarrow \texttt{Int},\\
  \phantom{....}\texttt{concat} & :    \texttt{Str} \times \texttt{Str} \rightarrow \texttt{Str},\\
  \phantom{....}\texttt{eqStr} & :     \texttt{Str} \times \texttt{Str} \rightarrow \texttt{Bool},\\
  \phantom{....}\texttt{mkPair} & :    \texttt{Int} \times \texttt{Int} \rightarrow \texttt{Pair},\\
  \phantom{....}\texttt{fst} & :       \texttt{Pair} \rightarrow \texttt{Int},\\
  \phantom{....}\texttt{snd} & :       \texttt{Pair} \rightarrow \texttt{Int},\\
  \phantom{....}\texttt{read} & :      \texttt{Path} \rightarrow \texttt{Str},\\
  \phantom{....}\texttt{join} & :      \texttt{Path} \times \texttt{Str} \rightarrow \texttt{Path},\\
  \phantom{....}\texttt{tmp} & :       \texttt{Int} \rightarrow \texttt{Path},\\
  \phantom{....}\texttt{parseDate} & : \texttt{Str} \rightarrow \texttt{Date},\\
  \phantom{....}\texttt{dateToInt} & : \texttt{Date} \rightarrow \texttt{Int},\\
  \phantom{....}\texttt{addDays} & :   \texttt{Date} \times \texttt{Int} \rightarrow \texttt{Date},\\
  \phantom{....}\texttt{isWeekend} & : \texttt{Date} \times \texttt{Bool},\\
  \phantom{....}\texttt{choose} & :    \texttt{Bool} \times \texttt{Int} \times \texttt{Int} \rightarrow \texttt{Int},\\
  \phantom{....}\texttt{mux} & :       \texttt{Bool} \times \texttt{Str} \times \texttt{Str} \rightarrow \texttt{Str}\\
  \end{align*}

  \section{Samples ($\sigma \leftsquigarrow \mathcal{L}(G_\Gamma') \cap \Sigma^{28}$)}
\vspace{0.3cm}
\begin{tiny}
  \texttt{fun f0 ( p1 : Pair ) : Float = i2f ( choose ( ( 1 == 1 ) , snd ( p1 ) , 1 ) )}\\
  \texttt{ fun f0 ( p1 : Bool , p2 : Date ) : Str = f0 ( ( ( p1 == p1 ) == true ) , p2 )}\\
  \texttt{ fun f0 ( p1 : Bool , p2 : Bool ) : Path = f0 ( if p2 \{ p1 \} else \{ true \} , p1 )}\\
  \texttt{ fun f0 ( p1 : Bool ) : Pair = f0 ( ( ( 1 * len ( i2s ( 1 ) ) ) < 1 ) )}\\
  \texttt{ fun f0 ( p1 : Int ) : Int = choose ( false , 1 , ( f0 ( 1 ) * f0 ( p1 ) ) )}\\
  \texttt{ fun f0 ( p1 : Int ) : Bool = ( s2i ( i2s ( choose ( true , p1 , 1 ) ) ) < p1 )}\\
  \texttt{ fun f0 ( ) : Bool = if true \{ true \} else \{ ( 1 < fst ( mkPair ( 1 , 1 ) ) ) \}}\\
  \texttt{ fun f0 ( ) : Bool = ( false == ( ( 1 == ( 1 + 1 ) ) == ( 1 < 1 ) ) )}\\
  \texttt{ fun f0 ( p1 : Str , p2 : Bool ) : Int = if eqStr ( p1 , p1 ) \{ 1 \} else \{ 1 \}}\\
  \texttt{ fun f0 ( p1 : Int , p2 : Int ) : Pair = f0 ( ( p2 * ( p2 + p2 ) ) , p1 )}\\
  \texttt{ fun f0 ( p1 : Int ) : Float = i2f ( ( p1 + len ( i2s ( ( 1 * p1 ) ) ) ) )}\\
  \texttt{ fun f0 ( p1 : Bool , p2 : Int ) : Str = f0 ( ( ( 1 == p2 ) == true ) , 1 )}\\
  \texttt{ fun f0 ( ) : Int = if if true \{ true \} else \{ false \} \{ 1 \} else \{ ( 1 * 1 ) \}}\\
  \texttt{ fun f0 ( p1 : Bool , p2 : Bool ) : Int = ( f0 ( false , ( p1 == false ) ) + 1 )}\\
  \texttt{ fun f0 ( p1 : Str ) : Int = ( 1 + s2i ( i2s ( ( f0 ( p1 ) * 1 ) ) ) )}\\
  \texttt{ fun f0 ( p1 : Bool , p2 : Pair ) : Bool = ( f0 ( p1 , p2 ) == ( p1 == false ) )}\\
  \texttt{ fun f0 ( p1 : Bool ) : Bool = ( false == ( f0 ( p1 ) == f0 ( f0 ( p1 ) ) ) )}\\
  \texttt{ fun f0 ( ) : Bool = ( true == ( i2s ( 1 ) == i2s ( s2i ( i2s ( 1 ) ) ) ) )}\\
  \texttt{ fun f0 ( ) : Int = if false \{ ( 1 * 1 ) \} else \{ ( 1 * ( 1 + 1 ) ) \}}\\
  \texttt{ fun f0 ( ) : Int = ( choose ( false , 1 , f0 ( ) ) + choose ( true , 1 , 1 ) )}\\
  \texttt{ fun f0 ( p1 : Bool , p2 : Bool ) : Int = f0 ( p2 , ( ( 1 + 1 ) == 1 ) )}\\
  \texttt{ fun f0 ( p1 : Str ) : Int = f0 ( i2s ( ( ( 1 + f0 ( p1 ) ) * 1 ) ) )}\\
  \texttt{ fun f0 ( ) : Bool = ( ( true == ( false == ( ( 1 == 1 ) == false ) ) ) == true )}\\
  \texttt{ fun f0 ( ) : Bool = ( if true \{ 1 \} else \{ ( 1 + 1 ) \} == ( 1 * 1 ) )}\\
  \texttt{ fun f0 ( ) : Int = ( ( ( 1 + f0 ( ) ) + s2i ( i2s ( 1 ) ) ) * 1 )}\\
  \texttt{ fun f0 ( p1 : Bool , p2 : Bool ) : Int = f0 ( ( true == ( p1 == p2 ) ) , p1 )}\\
  \texttt{ fun f0 ( p1 : Int , p2 : Int ) : Bool = f0 ( ( 1 + p1 ) , ( p2 + p2 ) )}\\
  \texttt{ fun f0 ( p1 : Path ) : Int = choose ( ( p1 == p1 ) , 1 , len ( i2s ( 1 ) ) )}\\
  \texttt{ fun f0 ( p1 : Float , p2 : Pair ) : Bool = f0 ( p1 , mkPair ( snd ( p2 ) , 1 ) )}\\
  \texttt{ fun f0 ( p1 : Int , p2 : Int ) : Path = f0 ( ( p1 + ( p1 + p1 ) ) , 1 )}\\
  \texttt{ fun f0 ( p1 : Pair , p2 : Path , p3 : Str ) : Int = s2i ( i2s ( s2i ( p3 ) ) )}\\
  \texttt{ fun f0 ( p1 : Int ) : Int = ( ( len ( i2s ( f0 ( 1 ) ) ) * 1 ) + 1 )}\\
  \texttt{ fun f0 ( ) : Bool = ( ( false == ( true == ( false == true ) ) ) == ( 1 == 1 ) )}\\
  \texttt{ fun f0 ( p1 : Pair ) : Int = choose ( ( 1 < s2i ( i2s ( 1 ) ) ) , 1 , 1 )}\\
  \texttt{ fun f0 ( p1 : Str , p2 : Str ) : Bool = ( ( false == true ) == f0 ( p2 , p1 ) )}\\
  \texttt{ fun f0 ( p1 : Pair , p2 : Path ) : Pair = f0 ( p1 , join ( p2 , i2s ( 1 ) ) )}\\
  \texttt{ fun f0 ( p1 : Int ) : Pair = mkPair ( p1 , ( ( 1 + p1 ) * ( p1 * p1 ) ) )}\\
  \texttt{ fun f0 ( p1 : Pair ) : Pair = mkPair ( 1 , ( 1 + ( 1 + ( 1 + 1 ) ) ) )}\\
  \texttt{ fun f0 ( p1 : Int ) : Int = ( f0 ( f0 ( 1 ) ) + ( f0 ( p1 ) + 1 ) )}\\
  \texttt{ fun f0 ( ) : Bool = ( ( true == ( 1 == 1 ) ) == ( ( 1 + 1 ) == 1 ) )}\\
  \texttt{ fun f0 ( p1 : Pair , p2 : Pair ) : Int = ( 1 * ( f0 ( p2 , p2 ) + 1 ) )}\\
  \texttt{ fun f0 ( p1 : Int , p2 : Str ) : Pair = mkPair ( ( ( p1 + p1 ) + 1 ) , p1 )}\\
  \texttt{ fun f0 ( p1 : Str ) : Path = tmp ( ( s2i ( i2s ( ( 1 + 1 ) ) ) + 1 ) )}\\
  \texttt{ fun f0 ( p1 : Int ) : Int = ( choose ( false , f0 ( 1 ) , p1 ) + f0 ( 1 ) )}\\
  \texttt{ fun f0 ( p1 : Date , p2 : Int ) : Int = f0 ( p1 , ( p2 + ( p2 * p2 ) ) )}\\
  \texttt{ fun f0 ( p1 : Int , p2 : Bool ) : Pair = mkPair ( ( p1 * ( p1 + p1 ) ) , p1 )}\\
  \texttt{ fun f0 ( p1 : Bool , p2 : Bool ) : Date = f0 ( ( true == p1 ) , ( p1 == true ) )}\\
  \texttt{ fun f0 ( p1 : Str ) : Path = tmp ( ( 1 * choose ( true , s2i ( p1 ) , 1 ) ) )}\\
  \texttt{ fun f0 ( p1 : Int , p2 : Int ) : Float = f0 ( ( ( p2 + p2 ) * p1 ) , p1 )}\\
  \texttt{ fun f0 ( p1 : Bool ) : Str = concat ( f0 ( true ) , i2s ( s2i ( i2s ( 1 ) ) ) )}\\
  \texttt{ fun f0 ( p1 : Int , p2 : Path ) : Str = i2s ( ( p1 + s2i ( read ( p2 ) ) ) )}\\
  \texttt{ fun f0 ( p1 : Date , p2 : Float ) : Bool = ( len ( read ( tmp ( 1 ) ) ) == 1 )}
\end{tiny}
  \clearpage
\end{document}